\documentclass[aps,pra,twocolumn,floatfix,superscriptaddress]{revtex4}

\usepackage{amssymb,amsmath,amstext}                
\usepackage{graphicx}                                               
\usepackage{epstopdf}                                               
\usepackage{xcolor}                                                     
\usepackage{bm}                                                        
\usepackage{appendix}                                              
\usepackage[utf8]{inputenc}
\usepackage{bbold}
\usepackage{bbm}
\usepackage{ulem}
\usepackage{braket}
\usepackage{latexsym}
\usepackage[colorlinks=true,citecolor=blue,linkcolor=magenta]{hyperref}

\newcommand{\vect}[1]{\mathbf{#1}}

\def\be{\begin{equation}}
\def\ee{\end{equation}}
\def\bea{\begin{eqnarray}}
\def\eea{\end{eqnarray}}

\def\bi{\begin{itemize}}
\def\ei{\end{itemize}}
\def\ben{\begin{enumerate}}
\def\een{\end{enumerate}}

\begin{document}

\title {Anderson Localization in Photonic Time Crystals}

\author{Karthik Subramaniam Eswaran}
\affiliation{Szkoła Doktorska Nauk Ścisłych i Przyrodniczych, Wydział Fizyki, Astronomii i Informatyki Stosowanej, Uniwersytet Jagiello\'nski, ulica Profesora Stanisława Łojasiewicza 11, PL-30-348 Kraków, Poland}
\affiliation{Instytut Fizyki Teoretycznej, Wydział Fizyki, Astronomii i Informatyki Stosowanej, Uniwersytet Jagiello\'nski, ulica Profesora Stanisława Łojasiewicza 11, PL-30-348 Kraków, Poland}
\author{Ali Emami Kopaei}
\affiliation{Szkoła Doktorska Nauk Ścisłych i Przyrodniczych, Wydział Fizyki, Astronomii i Informatyki Stosowanej, Uniwersytet Jagiello\'nski, ulica Profesora Stanisława Łojasiewicza 11, PL-30-348 Kraków, Poland}
\affiliation{Instytut Fizyki Teoretycznej, Wydział Fizyki, Astronomii i Informatyki Stosowanej, Uniwersytet Jagiello\'nski, ulica Profesora Stanisława Łojasiewicza 11, PL-30-348 Kraków, Poland}
\author{Krzysztof Sacha}
\affiliation{Instytut Fizyki Teoretycznej, Wydział Fizyki, Astronomii i Informatyki Stosowanej, Uniwersytet Jagiello\'nski, ulica Profesora Stanisława Łojasiewicza 11, PL-30-348 Kraków, Poland}

\begin{abstract}
Solutions of the wave equations for time-independent disordered media can exhibit Anderson localization where instead of wave propagation we observe their localization around different points in space. Photonic time crystals are spatially homogeneous media in which the refractive index changes periodically in time, leading to the formation of bands in the wave number domain. By analogy to Anderson localization in space, one might expect that the presence of temporal disorder in photonic time crystals would lead to Anderson localization in the time domain. Here, we show that indeed periodic modulations of the refractive index with the addition of temporal disorder lead to Anderson localization in time, where an electromagnetic field can emerge from the temporally modulated medium at a certain moment in time and then decay exponentially over time. Thus, we are dealing with a situation where, in a fluctuating three-dimensional medium, the birth and death of waves can occur, and the mechanism of this phenomenon corresponds to Anderson localization.
\end{abstract}

\date{\today}

\maketitle

Anderson localization is a phenomenon that appears in many different disordered systems \cite{Lagendijk2009}. Originally, Anderson localization involved the localization of a quantum particle in space due to the destructive interference of matter waves in the presence of a time-independent disordered potential \cite{Anderson1958}. Anderson localization can also be observed in momentum space, where it manifests as the quantum suppression of classical diffusion in phase space in systems that are classically chaotic \cite{Fishman:LocDynAnders:PRL82,Lemarie:Anderson3D:PRA09}. Recently, it has been shown that a quantum particle can also exhibit Anderson localization in time, where the probability of measuring the particle at a fixed position in space shows exponential localization around a certain moment in time if the perturbation of the particle contains temporal disorder \cite{Sacha15a,sacha16,delande17,SachaTC2020}. 

Photonic crystals in space are dielectric media in which the refractive index changes periodically in space \cite{Yablonovitch1993}. Solutions of the Maxwell equations in these media exhibit the characteristics of Bloch waves and we observe the formation of bands in the frequency domain. The research development in this field is vast \cite{Joannopoulos_Book}. If the medium is spatially homogeneous but the refractive index is periodically modulated in time, we are dealing with photonic time crystals \cite{Biancalana2007,Zurita2009,Galiffi2022,Kopaei2024a}. In this case, we also observe the formation of bands but in the wave number domain. The study of such systems falls under the field of time crystals, which has been developing dynamically in recent years \cite{Sacha2017rev,SachaTC2020,GuoBook2021,Zaletel2023}.

In media with time-independent disorder, Anderson localization can be observed when the solutions of wave equations show localization around different points in space. Research on photonic time crystals with temporal disorder and, more generally, on random time-varying media has been conducted \cite{Pendry1982,Sharabi2021,Carminati2021,Apffel2021,Zhou2024}. Among other things, it has been shown that the solutions of wave equations can exhibit exponential amplification over time in the presence of temporal disorder. However, none of these studies have shown that in photonic time crystals Anderson localization in the time domain is possible. Here, we demonstrate solutions whose amplitude grows exponentially over time up to a certain localization point, and then the solution decays exponentially over time and it all happens in the characteristic time scale called the localization time. 
This Anderson localization in time occurs simultaneously throughout the entire three-dimensional space, in contrast to Anderson localization in time in systems of massive particles driven resonantly \cite{Sacha15a,sacha16,delande17,SachaTC2020}. In the latter case, Anderson localization in time manifests as a wave packet propagating along the resonant trajectory, which, when observed at a fixed point in space close to the trajectory, exhibits an exponentially localized profile in time.

\begin{figure*}[t]
\centering
\includegraphics[width=1.\textwidth]{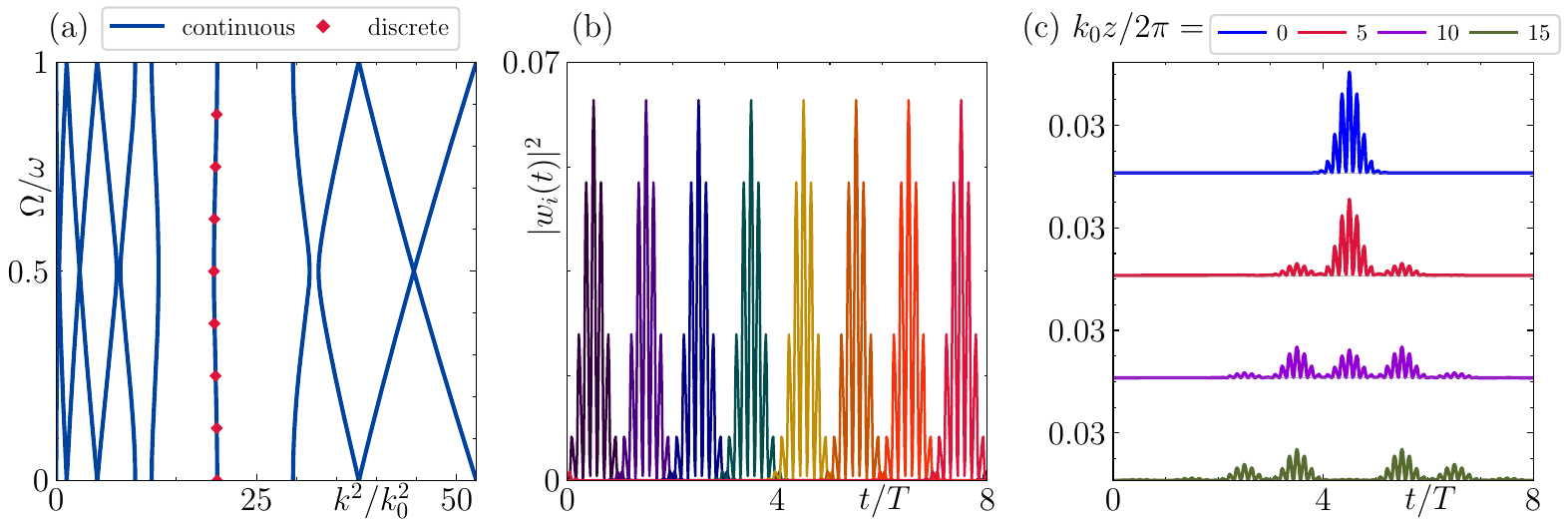}
\caption{
{\bf Disorder-free case.} (a) shows the band structure corresponding to the permittivity modulation as in Eq.~(\ref{epsilon}) for $\lambda_1=0.269$ and $\lambda_2=0.375$, where $\omega=2\pi/T$ and $k_0=2\pi/cT$. Red diamonds indicate quasi-frequencies $\Omega=(m-1)\omega/s$, with $m=1,\dots,s$ and $s=8$, chosen to obtain the Wannier states presented in (b). The Wannier states are localized in temporal sites of the photonic time crystal around $t_j=-T/2+jT$ with $j=1,\dots,s$. These Wannier states can be used to construct the basis for the photonic time crystal extending along the entire time axis. That is, any of the presented Wannier states can be translated by an integer multiple of $T$ to any temporal site of the photonic crystal. Top panel in (c) shows one of the Wannier states obtained to be localized in time when the electromagnetic field is observed at $z=0$. If we increase $z$ (see from top to bottom panels which correspond to $z=0$, 5, 10, 15, respectively), then the corresponding solution of the Maxwell equations shows that there is a transfer of the electromagnetic field from an initial temporal lattice site to neighboring temporal sites like in solid state physics where tunneling of a particle from a spatial lattice site to neighboring sites can be observed.}
\label{fig1}
\end{figure*}

In fiber optics, co-propagation of two waves, where the first wave is strong and reveals disorder in time and the other one is weak and localizes in the presence of the strong wave, was considered \cite{Agr-OL,Yazdani2024}. 
This effect arises due to the coupling between the waves in a nonlinear optical fiber and relies fundamentally on the one-dimensional geometry of the fiber. In contrast, here we consider a three-dimensional medium, where the electromagnetic field is governed by the linear Maxwell equations, enabling Anderson localization but also opening new possibilities for the realization of other condensed matter phases in the time dimension.

In the present Letter, we show that the description of photonic time crystals in a three-dimensional medium with weak temporal disorder can be reduced to the Anderson model (i.e., the tight-binding model with disorder) \cite{Anderson1958} and reveals Anderson localization in time. The tight-binding model is obtained by defining a basis of Wannier states, analogous to the standard description of spatial crystalline structures in condensed matter physics \cite{Wannier1937}. However, in our case, the Wannier states are not localized in space but in time around time instants separated by the period of the modulation of the medium. Consequently, the resulting Anderson localization described by the tight-binding model occurs in the time domain.

\begin{figure*}[t!]
\centering
\includegraphics[width=1.\textwidth]{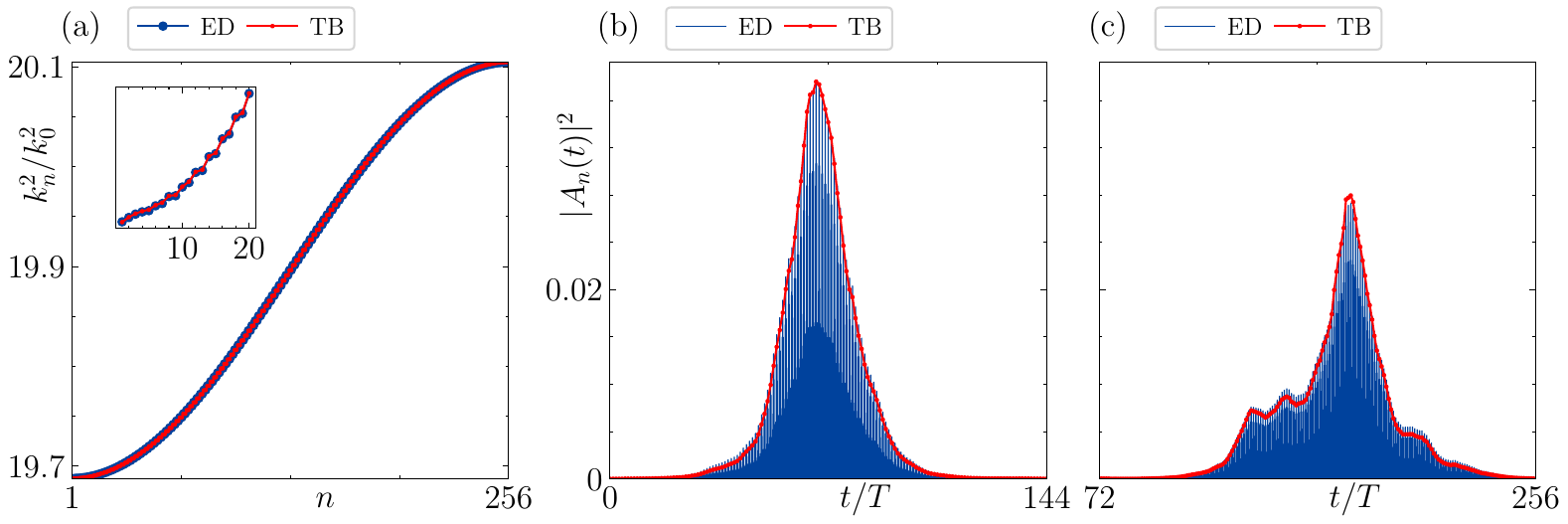}
\caption{{\bf Disordered case.} (a) presents eigenvalues $k_n^2$ obtained in the tight-binding model (red) for the temporal lattice of $s=256$ sites with disorder, Eq.~(\ref{am}), and the corresponding exact results (blue). A magnified fragment of the results is presented in the inset. In (b) and (c) comparison of two Anderson localized solutions $A_n(t)$ (with $n=1$ and $2$, respectively) obtained in the tight-binding (red) and exact (blue) description is presented. For clarity, the tight-binding results are presented as the modulus squares of overlaps with Wannier states. The temporal disorder is created by temporal fluctuations of the permittivity resulting in the Gaussian distribution of the on-site terms $E_j$ in (\ref{am}) with the standard deviation $\eta\chi$ where $\eta=1.2\times 10^{-4}$ and $\chi/c^2k_0^2=15.66$, see (\ref{ejdistr}). The other parameters are the same as in Fig.~\ref{fig1}.}
\label{fig2}
\end{figure*}

Consider a spatially homogeneous dielectric medium in which the relative permittivity changes periodically in time, $\varepsilon(t+T)=\varepsilon(t)$, and represent the electric and magnetic fields using the vector potential, $\vect{E} = -\partial_t \vect{A}$ and $\vect{B} = \nabla \times \vect{A}$, in the Coulomb gauge. Without loss of generality, we can restrict ourselves to solutions of the form $\vect{A}(\vect{r}, t) = \vect{e}_x A(t) e^{ikz}$ with $A(t)$ satisfying (see \cite{SM})
\be
{\cal H}A(t)\equiv\partial_t \left[\varepsilon(t) \partial_t A(t)\right]=-c^2k^2 A(t),
\label{H}
\ee
where $c$ is the speed of light in vacuum.
Since $\varepsilon(t)$ changes periodically in time, we can use the Floquet theorem \cite{Shirley1965,SachaTC2020} and look for $A(t)$ in the form $A(t)=A_\Omega(t)e^{-i\Omega t}$ where $A_\Omega(t+T)=A_\Omega(t)$ and $\Omega$ is the quasi-frequency of the electromagnetic wave, obtaining the equation
\be
{\cal H}_\Omega A_\Omega(t)\equiv(\partial_t-i\Omega) \left[\varepsilon(t) (\partial_t-i\Omega) A_\Omega(t)\right]=-c^2k^2 A_\Omega(t).
\label{Homega}
\ee
By fixing $\Omega$, this equation can be considered as an eigenvalue problem for the operator ${\cal H}_\Omega$ with the eigenvalue $-k^2$. For real values of $\Omega$, the operator ${\cal H}_\Omega$ is Hermitian, and the eigenvalues $k^2$ form bands which correspond to the bands in the wave number domain, $\pm k(\Omega)$. Figure~\ref{fig1} shows an example of the band structure for \be
\varepsilon(t)=1+\lambda_1\cos^2(7\pi t/T)+\lambda_2\cos^2(8\pi t/T),
\label{epsilon}
\ee
with $\lambda_1,\lambda_2\ge 0$. In photonic time crystals, the wave number $k$ lying in the gaps is also allowed. In this case, $\Omega$ takes complex values, and in a temporally modulated medium, an exponential growth in the intensity of the electromagnetic field over time is possible. In the following, we will restrict ourselves to solutions with wave numbers belonging to the bands, where $\Omega$ is real. More precisely, we will limit ourselves to a single band and investigate what happens to the solutions belonging to this band when fluctuations in permittivity modulations appear. 

In the Floquet approach, to describe bands in $k$, it is sufficient to restrict to the so-called Floquet zone \cite{SachaTC2020}, i.e., quasi-frequencies $\Omega$ within the interval of the modulation frequency of the permittivity, $\omega=2\pi/T$. This is fully analogous to the Brillouin zone in solid-state physics. So far, the changes of the permittivity are perfectly periodic. Shortly, we will turn on weak additional fluctuations of permittivity. We will assume that the additional perturbation is weak enough to significantly couple only states belonging to the same band, and the couplings between bands and with states in the gaps can be neglected. This description will allow us to identify and understand Anderson localization, and it will be confirmed in numerical solutions of the full Maxwell equations.

Let us assume that we want to reduce (\ref{H}) to a tight-binding model describing one of the bands visible in Fig.~\ref{fig1}. For this purpose, we need to introduce a basis of Wannier states, similar to what is done in a typical solid-state problem \cite{Wannier1937}. In our case, the Wannier states will be localized not at different positions in space, as in condensed matter, but at different moments in time. The Wannier states will be a superposition of solutions of (\ref{H}), i.e., $A_\Omega(t)e^{-i\Omega t}e^{ik(\Omega)z}$, corresponding to $k(\Omega)$ from the chosen band. Suppose that we want to obtain Wannier states whose localization in time will be observed at, e.g., $z=0$, and so the sought Wannier states can be obtained as a superposition of $A(t)=A_\Omega(t)e^{-i\Omega t}$. In condensed matter physics, Wannier states can be determined as eigenstates of the position operator in the Hilbert subspace restricted to a single band \cite{Kivelson1982}. In practice, an infinite crystal is not considered, but rather a finite crystal structure with periodic boundary conditions \cite{Dutta2015}. In our case, this means that we will diagonalize the operator $e^{i\omega t/s}$, where integer-valued $s$ determines the temporal size, $sT$, of the photonic time crystal \cite{SM}. Thus, we perform the diagonalization of the operator $e^{i\omega t/s}$ in the basis $A(t)=A_\Omega(t)e^{-i\Omega t}$ corresponding to a single band, where we restrict ourselves only to $\Omega = (m-1)\omega/s$ with integer $m$ from 1 to $s$. As a result, we obtain $s$ Wannier states, $w_j(t)= \sum_\Omega \alpha_{\Omega,j} A_\Omega(t)e^{-i\Omega t}$, 
localized at $t_j= -T/2 + jT$ where integer $j=1,\dots,s$ (Fig.~\ref{fig1}). If $s \gg 1$, the obtained Wannier states $w_j(t)$ can serve as a basis for a photonic time crystal with infinite temporal size, i.e. with $j$ going from $-\infty$ to $+\infty$, because $w_j(t)$ decays exponentially fast as $t$ moves away from the localization point $t_j$, and the fact that the Wannier states were obtained for $s < +\infty$ does not matter. Expanding solutions of the Maxwell equations in the Wannier basis, $A(t)=\sum_jc_jw_j(t)$, we can reduce (\ref{H}) to a tight-binding model, 
\be
Jc_{j+1}+J^*c_{j-1}=(c^2k^2-U)c_j,
\label{tb}
\ee
where the only important couplings are between nearest-neighbor temporal sites, $J=\int w^*_{j+1}(t){\cal H}w_j(t)dt$, and $U=\int w^*_j(t){\cal H}w_j(t)dt$ is the same for all sites \cite{SM}. Naturally, solving (\ref{tb}) leads to the solutions, $A(t)=A_\Omega(t) e^{-i\Omega t}$, from which we started.

Note that the localization of the obtained Wannier states is not observable at all $z$ but only at $z=0$. Interestingly, if we analyze what happens at other positions $z$, we come to the conclusion that $z$ plays the role of a fictitious time, similarly to its role in the optical Schrödinger equation \cite{marte1997paraxial}. If we start changing $z$ from 0, then $w_j(t,z) = \sum_\Omega \alpha_{\Omega,j} A_\Omega(t)e^{-i\Omega t}e^{ik(\Omega)z}$ will show that as $z$ increases, we observe transfer from the temporal site $t_j$ to neighboring sites (Fig.~\ref{fig1}), similar to what happens in condensed matter when we prepare a particle in a Wannier state and its evolution in time causes it to tunnel to neighboring sites. 

So far, the permittivity has been changing periodically in time with the period $T$. Now, let us turn on an additional perturbation to the permittivity, $\Delta\varepsilon(t)$, which fluctuates in time and is characterized by a correlation time not larger than $T$. For numerical simulations, we have to limit ourselves to a finite system with the finite number, $s$, of temporal lattice sites. By choosing $\Delta\varepsilon(t) =\eta\sqrt{2/s}\sum_{m=1}^s  \cos(m\omega t/s+\varphi_m)$ with $\varphi_m$ chosen randomly and uniformly in the interval $[0,2\pi)$, we arrive at the final tight-binding equations,
\be
Jc_{j+1}+J^*c_{j-1}+E_jc_j=(c^2k^2-U)c_j,
\label{am}
\ee
where $E_j=\int  w^*_j\partial_t[\Delta\varepsilon\partial_tw_j]dt$ which can be reduced to
\be
E_j\approx \eta\chi\sqrt{2/s}\sum _{m=1} ^{s} \cos\left( \varphi _m + m\omega t _j/s\right),
\label{ejdistr}
\ee
with $\chi=\int |\partial_tw_j|^2dt$.  Equation~(\ref{am}) is the standard Anderson model because $E_j$ are random numbers. Indeed, the central limit theorem guarantees that $E_j$
are random numbers chosen from the normal distribution with zero mean and standard deviation $\eta\chi$ \cite{SM}. If $\eta\chi$ is much smaller than the gaps between the considered band and the neighboring bands, the tight-binding model provides accurate predictions for the exact solutions of the Maxwell equations. Figure~\ref{fig2} presents the eigenvalues $k^2$ and examples of Anderson-localized eigenstates obtained by solving (\ref{am}) and the corresponding exact numerical solutions of the Maxwell equations. The latter were obtained using the Floquet theorem, leveraging the fact that in the presence of $\Delta\varepsilon(t)$ the system is still periodic in time but with a period $sT$, and finding solutions of an equation analogous to (\ref{Homega}) for the quasi-frequency $\Omega=0$.
\begin{figure}[t]
\centering
\includegraphics[width=1.\columnwidth]{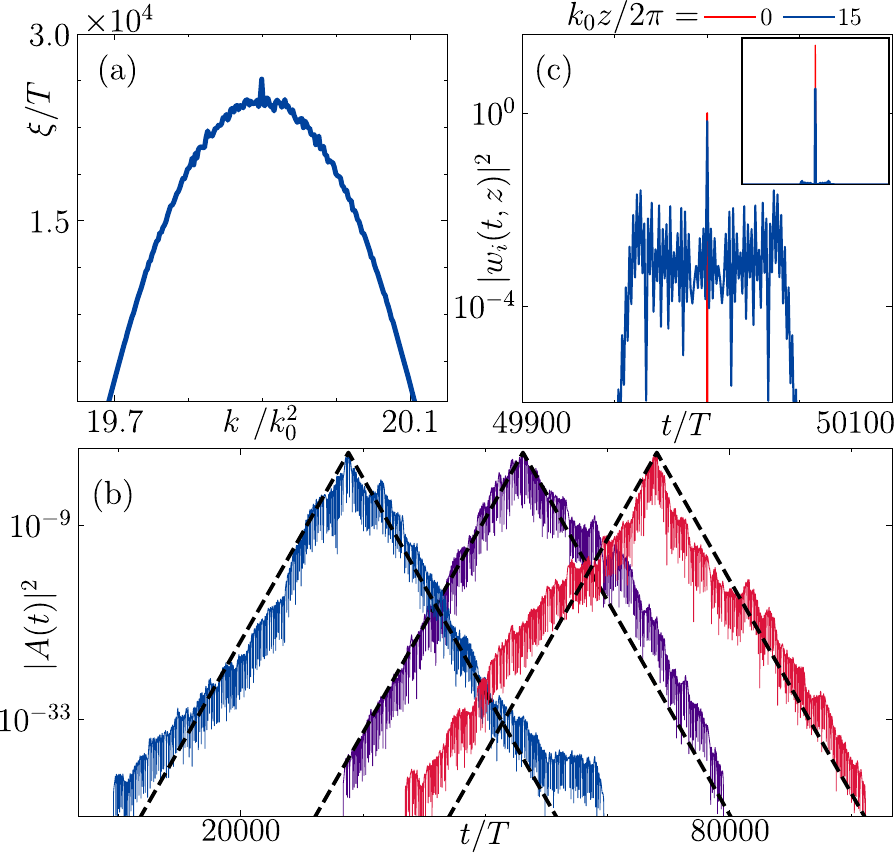}
\caption{{\bf Disordered case.} (a) presents localization time (analogue of localization length in standard Anderson localization in space) vs. wavenumber squared corresponding to the band analyzed in Figs.~\ref{fig1}-\ref{fig2}. The results have been obtained within the tight-binding model and describe the photonic time crystal with the infinite temporal size. (b) demonstrates localized eigenstates of the Anderson model (\ref{am}) in the vicinity of $k^2/k_0^2=19.692$, with an estimated localization length $\xi/T=493$ from transfer matrix calculations represented by the dashed lines. (c): if, instead of an eigensolution of the Maxwell equations, we choose, e.g., a solution which at $z=0$ reduces to a single Wannier state, then regardless of the location $z$ of the observation point, the electromagnetic field is localized in time due to the Anderson localization phenomenon. For this state, the inverse participation ratio, IPR=$\int|w_i(t,z)|^4dt=0.48/T$. The inset shows the same as the main plot but in the linear scale. Panel (c) should be compared with the similar panels in Fig.~\ref{fig1}(c) where in the absence of the temporal disorder, the electromagnetic field localized in time at  $z=0$ loses its localization properties if we change the observation point. For the corresponding state in Fig.~\ref{fig1}(c), IPR$=0.16/T$. In the disordered case, the IPR freezes for large $z$, while for the disorder-free system it continuously decreases as we increase $z$.}
\label{fig3}
\end{figure}
Having tested the tight-binding model, we can now consider a photonic time crystal with an infinite number of temporal sites with fluctuating permittivity, i.e., taking the limit $s\rightarrow\infty$. Applying the transfer matrix approach \cite{Derrida1984,SM} in the tight-binding model, we obtain how the localization time (analog of the localization length in a standard Anderson localization problem) changes with the wave number $k$ (Fig.~\ref{fig3}). We thus deal with solutions describing the electromagnetic field that grows exponentially in time and uniformly throughout the entire space, with a localization time dependent on the wave number, reaches its maximum at a certain point in time, and then decays exponentially over the same timescale.

Having eigenstates, $A_k(t)$, of the Anderson model (\ref{am}), we can combine them to form a state that, e.g., at $z=0$ will be localized in time just like a single Wannier state, i.e.
${\tilde A}(t,z=0)=w_j(t)=\sum_k\gamma_kA_k(t)$. If we analyze how this solution changes with increasing $z$ from 0, ${\tilde A}(t,z)=\sum_k\gamma_kA_k(t)e^{ikz}$, we will see that the transfer to neighboring sites of the photonic time crystal, which we observed previously (cf. Fig.~\ref{fig1}), eventually ceases because Anderson localization takes control over the system, see Fig.~\ref{fig3}(b). So once again, we see that $z$ can be treated as a fictitious time. Then, in full analogy to standard Anderson localization, we observe that an initially prepared localized state, which is not an eigenstate, begins to delocalize until Anderson localization takes over, and the wave packet spreading comes to a halt. Note that while each eigenstate of the Anderson model (\ref{am}) reveals exponential localization in time, a superposition of eigenstates with different localization times is not described by a single exponential function, as shown in Fig.~\ref{fig3}(c).

In summary, we considered photonic time crystals, i.e., systems in which permittivity is periodically modulated in time leading to the formation of bands in the wavenumber domain. The aim of our work was to demonstrate that in the presence of temporal disorder, Anderson localization is possible in photonic time crystals, which we observe in time rather than in space, as is typically the case. Previous studies have shown that temporal disorder can lead to the amplification of an electromagnetic field that grows exponentially over time and never ceases \cite{Pendry1982,Sharabi2021,Carminati2021,Apffel2021}. In this Letter, we show that with weak temporal disorder, Anderson localization can be observed in its full extent. In a medium fluctuating in time, it is possible to excite an electromagnetic field, which manifests as a burst of radiation that occurs uniformly throughout the entire space. This burst lasts for the localization time, which can be calculated by reducing the description to the standard Anderson model.

In the field of ordinary space crystals, the main question is whether we observe a regular arrangement of atoms in space at a fixed moment in time (the moment of detection). In the field of time crystals, we reverse the roles of time and space \cite{SachaTC2020}. We place a detector at a certain position in space and ask whether the probability of the detector clicking changes regularly over time. In this Letter, we also show that position in space within photonic time crystals can act as a fictitious time. 

Finally, we would like to point out that the phenomena we describe may be expected in many wave equations with temporal disorder. In a medium fluctuating in time, the birth and death of a wave may occur, reflecting the mechanism discovered by Philip Anderson \cite{Anderson1958}. Research on other condensed matter phenomena in photonic time crystals is in progress \cite{progress}.

We acknowledge the support of the
National Science Centre, Poland, via Project No. 2021/42/A/ST2/00017. The numerical computations in this work were supported in part by PL-Grid Infrastructure, Project No. PLG/2023/016644.

\bibliography{ref_tc_book}

\appendix
\section* {Supplemental material: Anderson Localization in Photonic Time Crystals}

In Section~\ref{A}, we derive Maxwell equations governing the system under time-modulated permittivity. By applying the Coulomb gauge and Floquet theorem, we reformulate the Maxwell equations as an eigenvalue problem for the wavenumber squared. Section~\ref{B} describes the tight-binding model for the system. Using Floquet solutions, we compute Wannier functions \cite{Wannier1937,Kivelson1982} which form a basis localized in time. In section \ref{C}, we model disorder in the time-domain by adding fluctuations in the permittivity modulation.  In the last section, we describe the transfer matrix calculation.

\subsection{Maxwell equations
\label{A}}

We consider Maxwell equations in a spatially homogeneous medium with the relative permittivity periodically modulated in time, $\varepsilon(t+T)=\varepsilon(t)$, representing the electric and magnetic fields in terms of the vector potential, $\vect{E} = -\partial_t \vect{A}$ and $\vect{B} = \nabla \times \vect{A}$, and assuming the Coulomb gauge, $\nabla\cdot \vect{A} = 0$. Without loss of generality we may propose, 
\be
\label{vpa}
\vect{A}(\vect r,t) = A(t) e^{ikz} \vect{\Hat{e}_x},
\ee
as an ansatz for TEM field modes 
which leads to the following wave equation 
\be
\label{mes}
{\cal H}A(t) \equiv \partial_t \left[\varepsilon(t) \partial_t A(t)\right]=-c^2k^2 A(t).
\ee
Because $\varepsilon(t)$ is periodic function of time, we can apply the Floquet theorem \cite{Shirley1965,SachaTC2020} and look for solutions of Eq.~(\ref{mes}) in the form
\be
A(t) = A_{\Omega}(t) e^{-i\Omega t},
\ee
where $A_\Omega(t+T)=A_\Omega(t)$ and $\Omega$ is a quasi-frequency which can take complex values but we focus on real $\Omega$ in the Letter. The resulting equation forms a well defined eigenvalue problem for the wavenumber squared, 
\be
{\cal H}_\Omega A_\Omega(t)\equiv(\partial_t-i\Omega) \left[\varepsilon(t) (\partial_t-i\Omega) A_\Omega(t)\right]=-c^2k^2 A_\Omega(t).
\label{Homega}
\ee
Since both $\varepsilon(t)$ and $A_\Omega(t)$ are $T$-periodic, for numerical calculations, it is convenient to switch to the frequency domain, $A_\Omega(t) = \sum _n A_\Omega ^n e^{in\omega t}$ and $\varepsilon(t) = \sum _n \varepsilon ^n e^{in\omega t}$ where $\omega=2\pi/T$, which transforms Eq.~(\ref{Homega}) into 
\be
\label{fdm}
\sum_{n'}(n\omega-\Omega)(n'\omega-\Omega)\varepsilon^{n-n'}A_\Omega ^{n'} = c^2k^2 A_\Omega ^n.
\ee
For real $\Omega$, this is an eigenequation for a Hermitian matrix that can be diagonalized by means of standard routines.

Equation (\ref{fdm}) is an eigenvalue problem solely parameterized by the quasi-frequency $\Omega$. In the Letter we focus on solutions corresponding to real values of $\Omega$. In the Floquet theory it is sufficient to consider $\Omega$ in a range of the frequency of the permittivity modulation. Changing $\Omega$, we observe that eigenvalues $k^2$ form bands (see Fig.~1 in the Letter). This situation corresponds to an infinite crystalline structure in time. If we restrict ourselves to discrete values of $\Omega_j =-\omega/2+(j-1)\omega/s,\ j = {1,\dots,s}$, we will obtain a finite crystal structure in time ($s$ temporal sites) which extends over the period $sT$ with periodic boundary conditions. In solid-state physics, this situation corresponds to a finite spatial crystalline structure with periodic boundary conditions in space. In the theoretical description, one is then able to numerically obtain the basis of Wannier states localized in the sites of the spatial crystal structure \cite{Dutta2015}. Since these states are exponentially localized, they do not sense that they were obtained for a finite crystal structure, and they can be used to describe an infinite spatial crystal. That is, having a single Wannier state, we can translate it by an integer multiple of the spatial lattice constant and cover the entire spatial axis from $-\infty$ to $+\infty$. We will use the same strategy to describe photonic time crystals here. For a finite crystalline structure in time, with $s$ temporal sites and periodic boundary conditions, we will obtain numerically Wannier state basis localized in the temporal sites. Then, we will introduce temporal fluctuations, which will satisfy periodic boundary conditions in the finite crystalline structure in time. As a result, we will obtain a finite time crystal with disorder. Importantly, since the disorder satisfies periodic boundary conditions with period $sT$, we can still use Floquet theory, but by looking for solutions with the period $sT$. This will allow us to test the tight-binding approximation, which we will ultimately apply to describe Anderson localization in infinite photonic time crystals.

\subsection{Tight binding description
\label{B}}

We consider a time-periodic modulation of the permittivity,
\be
\label{eps}
\varepsilon(t)=1+\lambda_1\cos^2\left(\frac{7\pi t}{T}\right)+\lambda_2\cos^2\left(\frac{8\pi t}{T}\right),
\ee
where $\lambda_1,\lambda_2>0$. Such a modulation can lead to a well-isolated band in the momentum space, which can be effectively described by a tight-binding model even in the presence of an additional weak temporal fluctuations of the permittivity. The latter we will introduce to realize temporal disorder in the photonic time crystal.

To derive the tight-binding model we first choose the Floquet solutions \begin{equation}
A_j(t)=A_{\Omega_j}(t)e^{-i\Omega_jt},
\end{equation} 
corresponding to a single band in the $k^2$ spectrum, with $\Omega_j=(j-1)\omega/s,\ j = {1,\dots,s}$ where $s$ is integer. In this subspace, we are looking for the Wannier basis of $s$ temporally localized states.
The most localized states in time should correspond to eigenstates of the {\it time operator} in the considered subspace. However, in the case of the periodic boundary conditions in time, it is more convenient to diagonalize a periodic function of such an operator like $\exp(i\omega t/s)$. Hence, we define the matrix $M$,
\be
M_{ij} =\int\limits_0^{sT} dt \;A_i^*(t)\;e^{i\omega t/s}\; A_j (t),
\ee
assuming that $\int_0^{sT}dt A_i^*A_j=\delta_{ij}$, which can be factorised by eigendecomposition,
\be
M = L ^\dagger \Lambda R,
\ee
where $\Lambda$ is the diagonal matrix containing the eigenvalues of $M$ and obtain the Wannier states 
\bea
w_i^{R(L)}(t)&=&\sum_{j=1}^{s}\alpha_{ji}^{R(L)}A_j(t),
\eea
where $\alpha^{R(L)} _{ij} = R(L)_{ij}$. Strictly speaking the Wannier states are biorthogonal, 
\be
\int\limits_0^{sT} dt \;w_i^{L*}(t) \;w^R_j(t) =\delta_{ij}.
\ee
However, if $s$ is not very small, the left and right Wannier states are indistinguishable numerically and we may use a single set of the Wannier states $w_j(t)\equiv w^R_j(t)\approx w_j^{L}(t)$. The matrix elements of $\cal H$, Eq.~(\ref{mes}), in the Wannier basis read, 
\begin{align}
    \label{tbh}
    {\cal H}_{ij} =\int\limits_0^{sT} dt\; w_i^{*}(t)\; \mathcal{H} \;w_j(t)= -c^2\sum_{l} \alpha ^{*}_{li} \; \alpha_{lj} \;k^2_l,
\end{align}
where $\alpha_{ij}\equiv \alpha_{ij}^R=\alpha_{ij}^L$.
Due to the $T$-periodicity of the permittivity, the diagonal elements are the same and we denote them by $U={\cal H}_{ii}$. We assume that the Wannier states are labeled so that $w_j(t)$ is localized around $t\approx -T/2+jT$, i.e. in the $j$-th temporal site. In the tight-binding approximation only the nearest neighbour couplings can be retained because $|{\cal H}_{ii+1}|\gg |{\cal H}_{ij}|$ with $j>1$. Modulus of all the nearest neigbour couplings ${\cal H}_{ii+1}$ is the same and we can even make their phases the same by suitable changes of the global phases of the Wannier states, $w_j(t)\rightarrow e^{i\phi_j}w_j(t)$. After such a transformation Eq.~(\ref{mes}), within the subspace corresponding to the chosen single band, takes the form of the tight-binding model 
\begin{equation}
    Jc_{j+1}+Jc_{j-1}=(c^2k^2-U)c_j,
\label{cleanTBM}
\end{equation}
where $J=-{\cal H}_{jj+1}$ and $c_i$'s are expansion coefficients of solutions of Eq.~(\ref{mes}) in the Wannier state basis.

\subsection{Modelling disorder in the time domain\label{C}}

In order to introduce temporal disorder in the photonic time crystal we add a weak fluctuating part,
\begin{equation}
    \Delta\varepsilon(t)=\frac{\eta}{\sqrt{2s}}\sum _{m=-s}^s e^{i\left(\varphi_m + m\omega t/s\right)},
\end{equation}
in the permittivity modulation so that the new permittivity $\varepsilon' (t) = \varepsilon (t) +\Delta\varepsilon(t)$, where the parameters $\varphi_m$ are chosen randomly from the uniform distribution over the interval $2\pi$. It results in additional onsite terms in the tight-binding model, 
\begin{equation}
    Jc_{j+1}+Jc_{j-1}+E_jc_j=(c^2k^2-U)c_j,
\label{disorderTBM}
\end{equation}
where 
\bea
E_j &=&\frac{\eta}{\sqrt{2s}}\sum _{m=-s} ^{s} \int\limits_0 ^{sT} dt \left|\partial_t w _j(t)\right|^2 e^{i \left(\varphi _m + m\omega t/s \right)} 
\cr && 
\label{Ei1}
\\
&\approx& \eta\chi\sqrt{\frac{2}{s}}\sum _{m=1} ^{s} \cos\left( \varphi _m + \frac{m\omega t _j}{s}\right),
\label{Ei2}
\eea
where 
\begin{equation}
    \chi=\int\limits_0 ^{sT} dt \left|\partial_t w _j(t)\right|^2.
\end{equation}
The final expression in Eq.~(\ref{Ei2}) follows from strong localization of the Wannier states around the centers of the temporal sites at $t_j=-T/2+jT$. By application of the central limit theorem, it is clear that $E_j$ should be normally distributed if $s$ is large. Numerically, we indeed find that values of $E_j$ generated with the help of (\ref{Ei1}) reproduce the normal distribution with zero mean and standard deviation $\eta\chi$. 

The tight-binding model (\ref{disorderTBM}), with onsite terms $E_j$ distributed randomly, is a standard Anderson model for localization in one-dimensional space. In the one-dimensional case, all eigenstates are exponentially localized regardless of how weak the disorder strength is but it requires a sufficiently large lattice (sufficiently large $s$ in our tight-binding model). If the standard deviation $\eta\chi$ of the distribution of $E_j$ is small and we start increasing the temporal size of the photonic time crystal from a small value of $s$, we first observe localization of eigenstates at the edges of the band because there the density of $k^2$ levels is the largest and it is the easiest to couple the unperturbed eigenstates by means of a weak disorder. When $s$ is sufficiently large, all eigenstates are exponentially localized. We are ultimately interested in a photonic time crystal with an infinite temporal size, and to determine localization times in the photonic time crystal, which correspond to localization lengths in the Anderson model (\ref{disorderTBM}), we will use a transfer matrix approach \cite{Derrida1984}, briefly described in the next section.

\subsection{Transfer matrix calculations of localization time}

In order to calculate localization lengths in the tight-binding model (\ref{disorderTBM}), we can use a transfer matrix approach \cite{Derrida1984}. It relies on propagation the system along the lattice starting with some initial sites. Equation~(\ref{disorderTBM}) can be rewritten in the transfer matrix form
\begin{align}
    \begin{pmatrix}
        c _{j+1} \\
        c_{j}
    \end{pmatrix}
    =
    \mathcal{T} _j
    \begin{pmatrix}
        c_{j} \\
        c_{j-1}
    \end{pmatrix},
\end{align}
where 
\begin{align}
    \mathcal{T}_j = 
    \begin{pmatrix}
        (c^2k^2 - U - E_j)/J & -1 \\
        1 & 0
    \end{pmatrix},
\end{align}
or equivalently, 
\begin{equation}
    R _{j+1} = \frac{c_{j+1}}{c_j} =  \frac{c^2k^2 - U - E_j}{J}- \frac{1}{R_j}.
\label{Ri}
\end{equation}
Starting from $R_1\ne 0$, the iteration of Eq.~(\ref{Ri}) allows us to obtain the localization length of the tight-binding model (\ref{disorderTBM}) corresponding to a given eigenvalue $k^2$ and consequently the localization time $\xi$ in the photonic time crystal,
\begin{equation}
    \xi=T\lim_{N\rightarrow\infty}\frac{N}{\sum_{j=1}^N\log|R_j|}.
\end{equation}

\end{document}